%% file: for_arxiv.tex
\documentclass[twoside, web, myside]{ieeecolor}
\usepackage{arxiv}
\usepackage{cite}
\usepackage{amsmath,amssymb,amsfonts}
\usepackage{algorithmic}
\usepackage{graphicx}
\usepackage{textcomp}
\usepackage{wrapfig,colortbl}
\usepackage{tabularx}
\definecolor{abstractbg}{rgb}{1,0.969,0.914}
\usepackage{placeins}

\newcommand{\rvec}{\mathbf{r}}
\newcommand{\kvec}{\hat{\mathbf{k}}}

\begin{document}
\title{On phase aberration estimation using common mid-angle speckle correlations}
\author{Naiara {Korta Martiartu} and Michael Jaeger
\thanks{This research was supported by the Swiss National Science Foundation under the Ambizione grant PZ00P2\_223375 to N.K.M. \textit{(Corresponding author: Naiara Korta Martiartu.)}}
\thanks{Naiara Korta Martiartu is with the Signal Processing Laboratory (LTS5), École Polytechnique Fédérale de Lausanne (EPFL), CH-1015 Lausanne,
Switzerland (e-mail: naiara.kortamartiartu@epfl.ch).}
\thanks{Michael Jaeger is with the Institute of Applied Physics, University of Bern, CH-3012 Bern,  Switzerland (e-mail: michael.jaeger@unibe.ch).}
\thanks{\textcopyright~2026 IEEE. Personal use of this material is permitted. Permission from IEEE must be obtained for all other uses, in any current or future media, including reprinting/republishing this material for advertising or promotional purposes, creating new collective works, for resale or redistribution to servers or lists, or reuse of any copyrighted component of this work in other works. DOI: 10.1109/TUSON.2026.3735057}}

\IEEEtitleabstractindextext{%
\begin{minipage}{\textwidth}\rightskip2em\leftskip\rightskip\bigskip
\begin{abstract}
Phase aberrations, despite degrading ultrasound images, also encode valuable information about the spatial distribution of the speed of sound in tissue. In pulse-echo ultrasound, we can quantify them by exploiting speckle correlations. Among existing strategies, correlations between steered acquisitions that share a common mid-angle have proven particularly effective for inferring the speed of sound. Their phases can be linearly related to the phase aberrations undergone by both the incident and reflected wavefronts. This relationship has so far been demonstrated only through geometric arguments based on point reflectors. Here, we develop a rigorous theoretical formalism that extends this relationship to the speckle regime, completing the previously established linear model and clarifying its underlying assumptions. More importantly, we build on this formalism to analyze correlation-phase fluctuations arising from aberration-induced speckle decorrelation. The analysis reveals that phase variance is governed by the relative loss of coherence, which increases approximately linearly with the square of the correlation phases. Local correlation-phase estimates therefore become increasingly uncertain as their magnitude grows. Experimental measurements in a uniform tissue-mimicking phantom show excellent agreement with the predicted variance. Beyond providing a theoretical basis for advancing speed-of-sound imaging, this formalism establishes the accuracy limit of common-mid-angle correlation phases, offering a benchmark for evaluating more advanced aberration-estimation techniques.

\end{abstract}

\bigskip
\end{minipage}}

\maketitle

\input{inputs/introduction_clean}
\input{inputs/theory_clean}

\FloatBarrier
\input{inputs/Experiment_clean}
\input{inputs/Conclusion_clean}

\input{inputs/Appendix_clean}


\bibliographystyle{IEEEtran}
\bibliography{references}

\end{document}

%% file: inputs/introduction_clean.tex
\section{Introduction}
\label{sec:introduction}

\IEEEPARstart{I}{n} pulse-echo ultrasound, we use a transducer array to insonify the tissue with a set of incident pressure waves. Variations in the acoustic impedance scatter the waves back, which are subsequently recorded by the same array. The measured pressure fields are then synthetically focused at each tissue position in both emission and reception to reconstruct a map of tissue reflectivity. This focusing operation forms the basis of conventional ultrasound imaging. It is typically implemented using delay-and-sum (DAS) beamforming, which assumes a uniform speed of sound to calculate the travel times of the incident and reflected waves. Tissues, however, often contain large-scale speed-of-sound heterogeneities that distort the ultrasound wavefronts, leading to discrepancies between the estimated and actual travel times. These phase errors, or phase aberrations, can severely degrade the resolution of ultrasound images and ultimately hinder their diagnostic interpretation~\cite{anderson_impact_2000}.

\begin{table*}[!t]
\arrayrulecolor{subsectioncolor}
\setlength{\arrayrulewidth}{1pt}
{\sffamily\bfseries\begin{tabular}{lp{6.75in}}\hline
\rowcolor{abstractbg}\multicolumn{2}{l}{\color{subsectioncolor}{\itshape
Highlights}{\Huge\strut}}\\
\rowcolor{abstractbg}$\bullet$ & A theoretical formalism is presented to relate common-mid-angle speckle correlations to the phase aberrations caused by speed-of-sound inhomogeneities. 
 \\
\rowcolor{abstractbg}$\bullet${\large\strut} & This formalism (i) completes and clarifies earlier models relating correlation phases to aberrations and (ii) predicts how aberrations affect the correlation-phase variance, verified experimentally.
\\
\rowcolor{abstractbg}$\bullet${\large\strut} & This fundamental understanding of speckle correlations is critical to advancing pulse-echo speed-of-sound imaging and assessing the accuracy limits of aberration-estimation techniques.\\[2em]\hline
\end{tabular}}
\setlength{\arrayrulewidth}{0.4pt}
\arrayrulecolor{black}
\end{table*}

Explicit estimation of phase aberrations provides a direct means of correcting their detrimental effects on ultrasound images. Once estimated, the phase errors can be subtracted from the initially assumed travel times to restore focus. Traditional techniques rely on the assumption that speed-of-sound inhomogeneities can be modeled as a thin phase screen located at the transducer face. The relative phase errors across the aperture, i.e., the phase screen, can then be estimated from the cross-correlation of signals received at neighboring elements~\cite{flax_phase-aberration_1988, odonnell_phase-aberration_1988}. This estimation is accurate in the presence of a bright, isolated point-like target (guide star); however, such targets are rare in biological tissue, where diffuse scatterers dominate. In this case, the accuracy depends on the degree of correlation between speckle signals~\cite{Mallart91, walker_speckle_1997}, and iterative focusing approaches are required to progressively reduce decorrelation and refine phase-aberration estimation~\cite{flax_phase-aberration_1988, Montaldo_FocusSpeckle}. Nevertheless, the near-field phase-screen model assumes that aberrations are spatially invariant over the entire medium (isoplanatism), meaning that a single phase-error profile, or phase-aberration law, can be applied to the beamformer to compensate for their effects. While this simplification may be reasonable for localized aberrators such as a superficial adipose layer, speed-of-sound inhomogeneities generally occur throughout the tissue, causing aberrations that remain invariant only within small regions (isoplanatic patch). Therefore, aberrations should ideally be estimated and corrected locally.

The common approach to local phase-aberration estimation is to exploit the correlations between signals acquired from different emissions, receptions, or both~\cite{jaeger_full_2015, chau_locally_2019, Lambert2020, lambert_ultrasound_2022-1, Bendjador2020}. The locally adaptive phase aberration correction (LAPAC) method~\cite{chau_locally_2019}, for example, extracts phase-aberration laws by analyzing correlations among time-delayed waveforms recorded from single-element emissions. Although such emissions benefit from spatial reciprocity, allowing the same correction to be applied on both transmit and receive, they inherently suffer from a low signal-to-noise ratio and are thus rarely used in practice. Ultrasound matrix imaging, which decouples transmit and receive focal spots during beamforming, offers a more general framework for quantifying aberrations under arbitrary acquisition schemes~\cite{lambert_reflection_2020, Lambert2020}. With simple matrix operations, the transmit (or receive) focal spots can be projected, for example, into the far-field to synthesize plane-wave emissions, while maintaining the receive (or transmit) side focused. This enables correlations between different emissions (or receptions) to be analyzed directly on the focused basis, which is more practical than operating on the raw waveforms. Because aberrations also affect the focused side, the estimated aberration laws inevitably mix contributions from both incident and reflected waves. To address this, an iterative procedure is employed, alternating estimation and correction between transmit and receive, to progressively improve the focusing quality and refine the aberration laws~\cite{lambert_ultrasound_2022-1, bureau_three-dimensional_2023}. Nevertheless, the interpretation of the final estimates is challenging, as the iterative correction cannot address axial aberrations~\cite{bureau_reflection_2024}. This limitation arises from the fundamental inability of correlations to constrain absolute phase aberrations. Knowledge of tissue speed of sound appears essential to fully correct their effects on ultrasound images~\cite{jaeger_full_2015, bureau_reflection_2024}.

In the context of speed-of-sound imaging, alternative approaches have been proposed to jointly estimate phase aberrations in both transmit and receive~\cite{stahli2019forward, simson_ultrasound_2025}, recognizing the fact that their individual contributions are difficult to disentangle. In particular, computed ultrasound tomography in echo-mode (CUTE) estimates phase aberrations by exploiting correlations between steered acquisitions that share a common mid-angle~\cite {stahli2019forward, jaeger2022pulseecho, NatureCUTE2022}. Using entirely geometrical arguments based on point-like reflectors, Staehli et al.~\cite{stahli2019forward} demonstrated that the phases of these correlations are linearly related to the phase errors accumulated along both transmit and receive paths during beamforming. Nevertheless, a more rigorous theoretical formalism is still lacking to understand the assumptions underlying this relationship and to characterize the stochastic nature of correlation phases in speckle. Establishing such a formalism is the primary goal of this paper. 

We organize the paper into two main parts. The first part (Section~\ref{sec:theory}) derives a closed-form expression for the expectation of common-mid-angle correlations in the presence of diffuse scatterers, with particular emphasis on the correlation phases. The second part (Section~\ref{sec:fluctuations}) then analyzes speckle-induced fluctuations in correlation-phase measurements and shows that their variance increases with the phase magnitude due to the coherence loss induced by aberrations. This prediction is validated experimentally in Section~\ref{sec:experiment} using a homogeneous tissue-mimicking phantom. Concluding remarks and the significance of this work are presented in Section~\ref{sec:conclusion}.

\begin{table*}[t!]
\caption{Glossary of mathematical notation.}
\label{tab:notation}
\centering
\renewcommand{\arraystretch}{1.15}
\setlength{\tabcolsep}{5pt}
\begin{tabularx}{\textwidth}{
    >{\centering\arraybackslash}p{0.10\textwidth}
    >{\raggedright\arraybackslash}X
    >{\centering\arraybackslash}p{0.10\textwidth}
    >{\raggedright\arraybackslash}X}
\hline
\textbf{Symbol} & \textbf{Definition} &
\textbf{Symbol} & \textbf{Definition} \\
\hline
$\mathbf r$ &
Position vector in tissue &
$\mathbf r_0$ &
Position vector in the reconstructed image \\
$r_\text{m}\ \text{or}\ r$ &
Coordinate of $\mathbf r$ along the mid-angle direction $\hat{\mathbf k}_\text{m}$; the subscript $\text{m}$ is dropped from Section~\ref{sec:CMA_correlations} onward  &
$r_{0,\text{m}}\ \text{or}\ r_{0}$ &
Coordinate of $\mathbf r_0$ along the mid-angle direction $\hat{\mathbf k}_\text{m}$ \\
$\delta r$ &
Relative distance between $r$ and $r_0$ &
$r_\text{c}$ &
Center position of the correlation kernel \\
$\hat{\mathbf k}_{\text{in}},
 \hat{\mathbf k}_{\text{out}}$ &
Transmit and receive propagation directions &
$\hat{\mathbf k}_\text{m},\hat{\mathbf k}_\text{p}$ &
Directions parallel and perpendicular to the mid-angle \\
$\varphi_{\text{in}},
 \varphi_{\text{out}}$ &
Transmit and receive steering angles &
$\varphi_\text{m},\varphi_\text{d}$ &
Mid-angle and angle difference, defined as:
$\varphi_\text{m}=(\varphi_{\text{in}}+\varphi_{\text{out}})/2$,
$\varphi_\text{d}=(\varphi_{\text{in}}-\varphi_{\text{out}})/2$ \\
$\chi(\mathbf r)$ &
Tissue scattering function &
$\widetilde{\chi}(r_\text{m})\ \text{or}\ \widetilde{\chi}(r)$ &
Projection of $\chi$ along $\hat{\mathbf k}_\text{p}$ at position $r_\text{m}$ or $r$ \\
$I(r_0)$ &
Reflectivity image reconstructed with single transmit-receive plane waves & 
$I_1, I_2$ & Two images reconstructed with common $\varphi_\text{m}$ but different $\varphi_{\text{d},1}$ and $\varphi_{\text{d},2}$.
 \\
$t( r)$ &
True total travel time of transmit-receive plane waves for a reflector at $r$ &
$t_0( r_0)$ &
Total travel time assuming a uniform speed of sound $c_0$, for a reflector at $r_0$ \\
$\tau(r_0,r)$ &
Travel time difference between $t_0( r_0)$ and $t( r)$  &
$\Delta t(r_0)$ &
Travel-time error or phase aberration at position $r_0$ \\
$A_i$ & Point-spread function of the reconstructed image $I_i$ &
$\mu_{\text{A}_i},\sigma_{\text{A}_i}$ &
Center and spatial width of $A_i$, defined in~\eqref{eq:MeanSDA} \\
$B=A_1A_2$ &
Product of two point-spread functions, corresponding to the images $I_1$ and $I_2$ &
$\mu_\text{B},\sigma_\text{B}$ &
Center and spatial width of $B$, defined in~\eqref{eq:MeanSDB} \\
$B_0$ &
Peak value of $B$; measure of local coherence between
$I_1$ and $I_2$, defined in~\eqref{eq:B0_main} &
$\beta$ &
Angle-dependent factor accounting for the mismatch between
the widths of $A_1$ and $A_2$, defined in~\eqref{eq:beta} \\
$\left\langle C \right\rangle$ &
Expected value of the common-mid-angle correlation &
$\widehat C$ &
Estimator of the common-mid-angle correlation from a single realization  \\
$\phi$ &
Correlation phase &
$\sigma_\phi^2$ &
Variance of the correlation phase \\
$\alpha$ &
Angle-dependent factor in the relationship between correlation phases and phase aberrations, defined in~\eqref{eq:IntegrandPhaseConstant}
 &
$\overline{\sigma}_\text{A}$ &
Effective average between $\sigma_{\text{A}_1}$ and $\sigma_{\text{A}_2}$, defined in~\eqref{eq:avgPSFwidth} \\
$L$ &
Size of the correlation kernel &
$N=L/\overline{\sigma}_\text{A}$ &
Number of resolution cells within the kernel \\
\hline
\end{tabularx}
\end{table*}

%% file: inputs/theory_clean.tex
\section{Theory of common mid-angle correlations}
\label{sec:theory}

Common-mid-angle correlations are typically measured using steered beams. These are synthesized in post-processing by focusing in transmit and receive over a small angular aperture~\cite{stahli2019forward}. The limited aperture yields relatively broad beams whose wavefronts can be treated as locally planar. In this section, we therefore consider plane waves in both emission and reception for analytical tractability. We first provide a theoretical model for images reconstructed from transmit-receive plane-wave measurements (Section~\ref{sec:Transmit-receivePW}) and then derive the expected value of their common-mid-angle correlations (Section~\ref{sec:CMA_correlations}). We finally focus on the resulting correlation phase and compare it against a previously established model (Section~\ref{sec:new_insights}), before discussing the practical estimation of common-mid-angle correlations and its underlying assumptions (Section~\ref{sec:ensavg}). Table~\ref{tab:notation}
summarizes the main mathematical notation used throughout this work.

\subsection{Images with Transmit-Receive Plane Waves}
\label{sec:Transmit-receivePW}
We consider an experiment in which a plane wave is emitted into the tissue along the propagation direction ${\kvec_\text{in} = (\sin\varphi_\text{in}, \cos\varphi_\text{in})}$, and the response is measured along the direction ${\kvec_\text{out} = (\sin\varphi_\text{out}, \cos\varphi_\text{out})}$ by applying similar time delays as in emission~\cite{Baptiste23}. We assume that the emitted waveform is a Gaussian-modulated sinusoidal pulse,
\begin{equation}
p_e(t) = A\left(t\right) e^{i\omega_c t} = e^{-t^2/2\sigma^2} e^{i\omega_c t},
 \label{eq:Pulse}   
\end{equation}
where $\omega_c$ is the angular center frequency, and the temporal width $\sigma$ of the Gaussian envelope~$A$ controls the bandwidth. 

\begin{figure}[t]
\centering
\includegraphics[width=1\columnwidth]{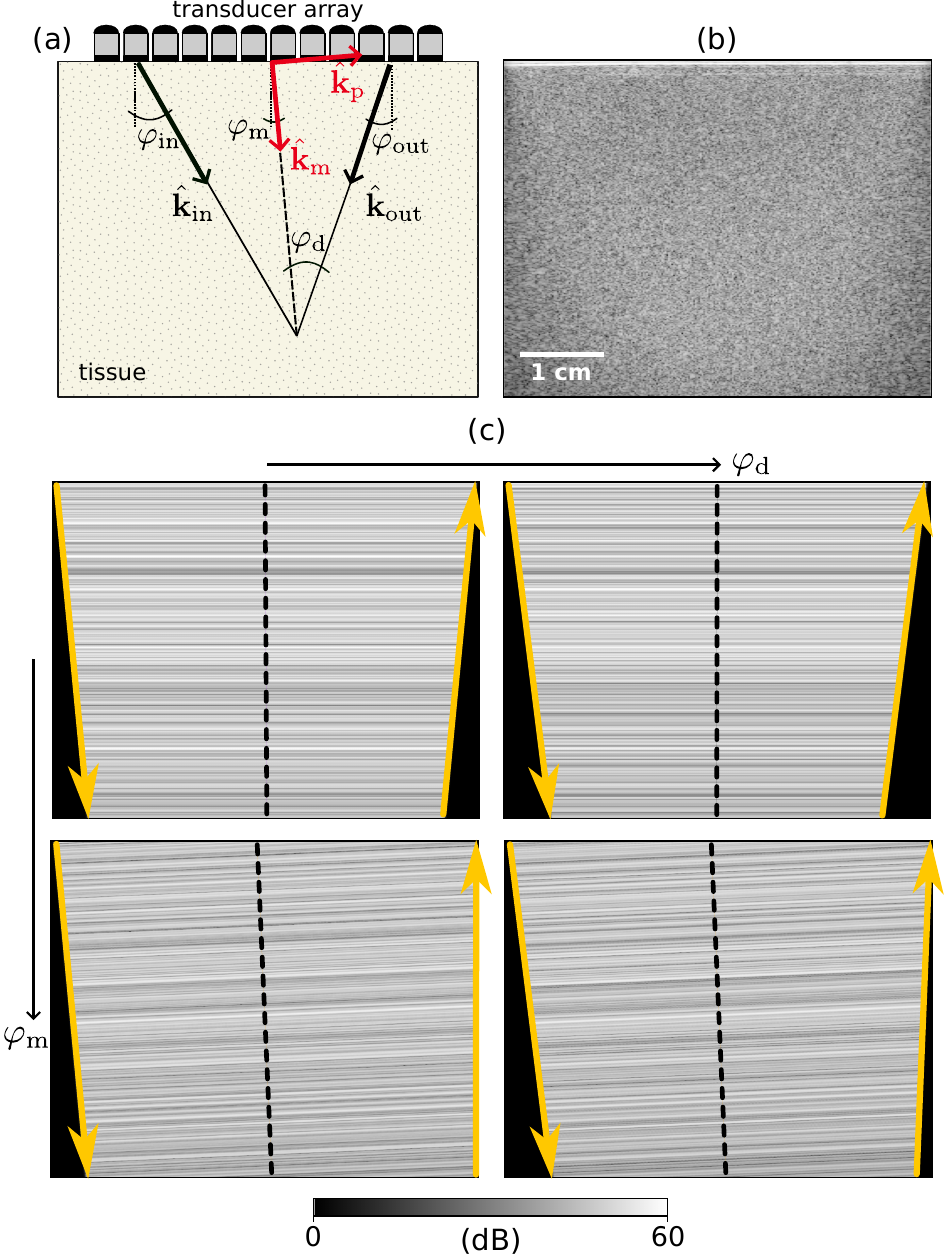}
\caption{Geometry and experimental illustration of common mid-angle correlations. (a) Acquisition geometry defining the mid-angle $\varphi_\text{m}$ and angle difference $\varphi_\text{d}$, together with the coordinate system (red) used in the analysis of common-mid-angle correlations [see~\eqref{eq:coordinate_change}]. (b) Coherently compounded reflectivity image reconstructed from data acquired in a uniform tissue-mimicking phantom (see Section~\ref{sec:experiment}), shown for reference. (c) Reflectivity images reconstructed from single transmit-receive plane-wave measurements acquired in the same phantom. Each row corresponds to a transmit-receive pair (yellow arrows) with the same mid-angle $\varphi_\text{m}$ but different angle difference $\varphi_\text{d}$; dashed lines indicate the corresponding mid-angle orientation. The images are spatially invariant along the direction perpendicular to $\kvec_\text{m}$; therefore, images acquired with the same $\varphi_\text{m}$ remain well correlated. }
\label{fig:PlaneWave_images}
\end{figure}

Let $\chi({\rvec})$ denote the scattering function of the tissue in the domain ${\Omega \subset \mathbb{R}^2}$. In the single-scattering regime, the image reconstructed through time-delay compensation of the measured response can be expressed as
\begin{equation}
I(\rvec{_0}) = \int_{\Omega} \chi({\rvec}) A\left(t_0(\rvec_{0}) - t({\rvec})\right) e^{i\omega_c\left(t_0(\rvec_{0}) - t({\rvec})\right)}d{\rvec}.
 \label{eq:Image}   
\end{equation}
The subscript 0 distinguishes the image position $\rvec_0$ from the tissue position $\rvec$, and $t$ is the total travel time of the recorded waves, while $t_0$ is computed assuming a uniform speed of sound $c_0$: 
\begin{equation}
t_0(\rvec_{0}) = \frac{1}{c_0}\left(\kvec_\text{in} + \kvec_\text{out}\right)\cdot\rvec_{0} = \frac{2}{c_0}\cos\left(\frac{\varphi_\text{in} - \varphi_\text{out}}{2}\right)\kvec_\text{m}\cdot\rvec_{0},
 \label{eq:tt_law}   
\end{equation}
with
\begin{equation}
\kvec_\text{m} = \left(\sin\left(\frac{\varphi_\text{in} + \varphi_\text{out}}{2}\right),~\cos\left(\frac{\varphi_\text{in} + \varphi_\text{out}}{2}\right)\right).
 \label{eq:kmid}   
\end{equation}
We define the mid-angle ${\varphi_\text{m}=(\varphi_\text{in} + \varphi_\text{out})/2}$ and the angle difference ${\varphi_\text{d}=(\varphi_\text{in} - \varphi_\text{out})/2}$, for convenience (Fig.~\ref{fig:PlaneWave_images}(a)). 

Equation~\eqref{eq:tt_law} shows that, for this type of measurement, the isochrones are straight lines perpendicular to the direction defined by the mid-angle between the transmit and receive steering angles. Therefore, 
we define a new coordinate system  aligned with the mid-angle direction $\kvec_\text{m}$, in which
\begin{equation}
{\rvec_0 = r_{0,\text{m}}\,\kvec_\text{m} + r_{0,\text{p}}\,\kvec_\text{p}} \qquad {\rvec = r_\text{m}\,\kvec_\text{m} + r_\text{p}\,\kvec_\text{p}},
\label{eq:coordinate_change}
\end{equation}
with $r_{0,\text{m}}, r_\text{m} \in \Omega_\text{m}$ and $ r_{0,\text{p}}, r_\text{p} \in \Omega_\text{p}$,
such that ${\Omega = \Omega_\text{m} \times \Omega_\text{p}}$. The vector $\kvec_\text{p}$ is the direction perpendicular to $\kvec_\text{m}$ (Fig.~\ref{fig:PlaneWave_images}(a)). In this coordinate system, the image $I(\rvec_{0})$ reduces to
\begin{equation}
I(r_{{0},\text{m}}{; \varphi_\text{d}}) =  \!\int_{\Omega_\text{m}} \! \! \tilde{\chi}({r}_\text{m})A\left(\tau(r_{{0}, \text{m}},{r}_\text{m} )\right)  e^{i\omega_c\tau(r_{{0}, \text{m}},{r}_\text{m} )}d{r}_\text{m},
 \label{eq:Image2}   
\end{equation}
with 
\begin{equation}
\tau(r_{{ 0}, \text{m}},{r}_\text{m}{; \varphi_\text{d}}) = t_0(r_{{0}, \text{m}}{; \varphi_\text{d}}) - t({r}_\text{m}{; \varphi_\text{d}})
\label{eq:tau_general}
 \end{equation}
and
\begin{equation}
t_0(r_{{0}, \text{m}}{; \varphi_\text{d}}) = \frac{2}{c_0}\cos\varphi_\text{d}\,r_{{0}, \text{m}}.
 \label{eq:tt_law2}
 \end{equation}
The scattering function $\tilde{\chi}(r_\text{m}) = \int_{\Omega_\text{p}}\chi(r_\text{m}, r_\text{p}) dr_\text{p}$ represents the average tissue reflectivity along the direction $\kvec_\text{p}$. 

Equations~\eqref{eq:Image2}--\eqref{eq:tt_law2} show that the point-spread function $A$ (PSF), and therefore the image $I(\rvec_{0})$, varies only along the mid-angle direction (Fig.~\ref{fig:PlaneWave_images}(c)). As a result, the images obtained with transmit-receive plane waves sharing a common mid-angle $\varphi_\text{m}$ encode similar spatial frequencies of $\chi({\rvec})$ and are strongly correlated. An explicit expression for these correlations is derived in the next subsection, where the subscript $\text{m}$, denoting the coordinate along the mid-angle direction, is omitted to simplify the notation.

\subsection{Correlation of Images with Common Mid-Angle}
\label{sec:CMA_correlations}
\mbox{}\\[-1em]
We consider two recordings corresponding to the angle combinations ${(\kvec^1_\text{in}, \kvec^1_\text{out})}$ and ${(\kvec^2_\text{in}, \kvec^2_\text{out})}$, where ${\kvec^1_\text{m} = \kvec^2_\text{m}}$. That is, the two acquisitions share a common mid-angle ${\varphi_{\text{m},1} = \varphi_{\text{m},2}}$ but have different angle differences ${\varphi_{\text{d}, 1} \neq \varphi_{\text{d}, 2}}$. Let ${I_1(r_0):=I(r_0; \varphi_{\text{d},1})}$ and ${I_2(r_0):=I(r_0; \varphi_{\text{d},2})}$ denote their respective images obtained as in~\eqref{eq:Image2}, and we use
the same subscript convention for $\tau$, $t_0$, and $t$ in what follows. Our goal is to derive the expected value of the correlation between $I_1$ and $I_2$ at location $r_0$, i.e.,
\begin{equation}
{\left\langle C({r_0}) \right\rangle = \left\langle I_1(r_{0})I^*_2(r_{0}) \right\rangle} , 
 \label{eq:Xcorr2}   
\end{equation}
where the brackets denote ensemble averaging over realizations of disorder.

In a first approximation, we model the tissue as a random distribution of unresolved scatterers, leading to the speckle pattern commonly observed in ultrasound images. Tissue scatterers are therefore assumed to be spatially uncorrelated, i.e., ${\langle
\tilde{\chi}({{r}})\tilde{\chi}({{r^\prime}}) \rangle = \tilde{X}\delta({{r - r^\prime}})}$, with $\delta$ the Dirac distribution and $\tilde{X} = \langle|\tilde{\chi}|^2\rangle$~\cite{Mallart91}. Under this assumption and using the expression in~\eqref{eq:Image2}, the correlation becomes
\begin{equation}
{\langle C (r_0) \rangle}= \tilde{X} \int_{\Omega}  B\left(r{_0, r} \right) e^{i\omega_c\left(\tau_1(r{_0, r}) - \tau_2(r{_0, r})\right)} d{r},
 \label{eq:Integrand1}   
\end{equation}
with
\begin{equation}
B\left(r{_0, r} \right) = A\left(\tau_1(r{_0, r} )\right) A\left(\tau_2(r{_0, r} )\right).
 \label{eq:IntegrandAmp} 
 \end{equation}
 Since the PSFs ${A_i := A\left(\tau_i(r{_0, r} )\right)}$ are Gaussian functions, the function $B$ is also a Gaussian, and we can leverage the symmetry around its mean to solve the integral in~\eqref{eq:Integrand1}.
 The mean $\mu_\text{B}$ and standard deviation $\sigma_\text{B}$ of $B$ can be analytically computed from the means $\mu_{\text{A}_i}$ and standard deviations $\sigma_{\text{A}_i}$ of $A_i$.
 To derive their expression, we write the tissue location $r$ as a deviation from the image location: ${r = r_0 + \delta r}$. This allows us to approximate the true travel times as
 \begin{equation}
t({r}) {=} t(r{_{0}}) + {\left[t(r_0+\delta r) -t(r_0)\right]} \approx t(r_{_0}) +  t_0(\delta {r}),
 \label{eq:LinearApp} 
 \end{equation}
where we assume that, for small $\delta {r}$, the difference between the true and assumed travel times is negligible. We will discuss the implications of this approximation later in this section. 

\begin{figure*}[ht!]
\centering
\includegraphics[width=1\textwidth]{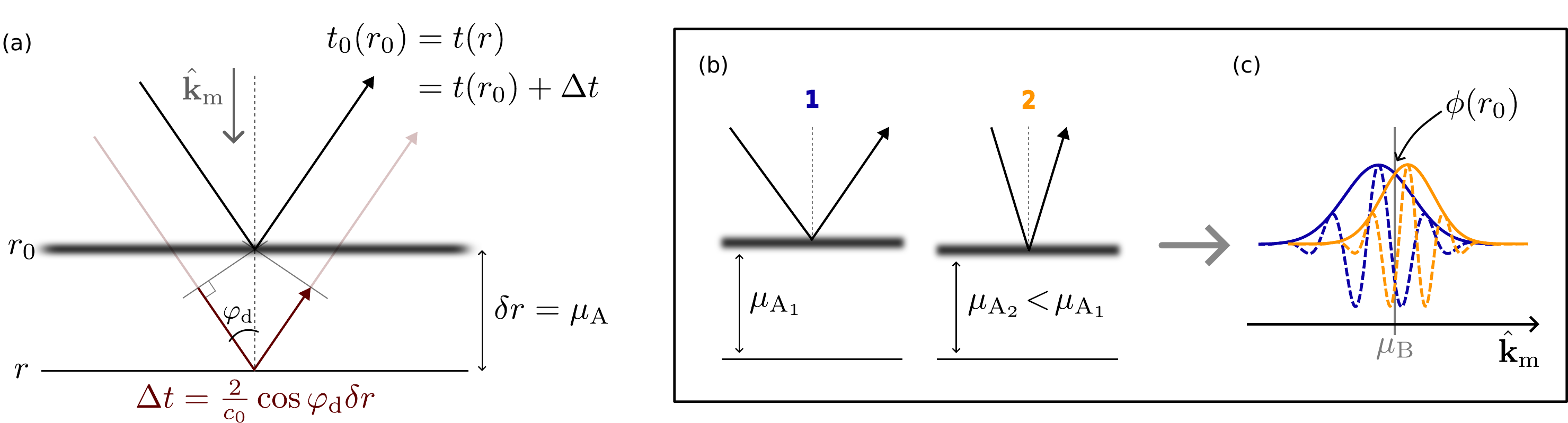}
\caption{Schematic of phase-aberration estimation using common mid-angle correlations. (a) The echo from a true reflector at $r$ is reconstructed at the image position $r_0$ satisfying $t_0(r_{0}) = t({r})$. When the true travel time $t$ and the assumed travel time $t_0$ differ at the same position, the reconstructed echo is shifted away from its true position along the mid-angle direction $\kvec_\text{m}$. The shift $\delta {r}$ depends on the travel-time error $\Delta t$ (phase aberration) and on the angle difference $\varphi_\text{d}$ between the transmit and receive directions. (b) Acquisitions sharing the same mid-angle but having different $\varphi_\text{d}$ yield different echo displacements. (c) The correlation phase $\phi(r_{_0})$ corresponds to the phase difference of the reconstructed echoes, evaluated at the peak of the product of their point-spread functions.}
\label{fig:Illustration}
\end{figure*}

By inserting~\eqref{eq:LinearApp} into~\eqref{eq:tau_general}, the travel time differences $\tau_i$ between positions ${r}$ and $r_{0}$ can be written in terms of the differences $\Delta t_i(r_{0}) = t_{0,i}(r_{0}) - t_i(r_{0})$ at the image location $r_{0}$:
\begin{equation}
\tau_i(r{_0, r}) \approx \Delta t_i(r_{0}) - \frac{2}{c_0}\cos\varphi_{\text{d},i}\delta {r}.
 \label{eq:TauApp} 
 \end{equation}
$\Delta t_i(r_{0})$ corresponds to what is commonly referred to as time-delay errors or phase aberrations. In contrast, $\tau_i$ describes how the phase of the reconstructed echo varies spatially. For a reflector located at $r$, the reconstructed echo is centered at the image location $r_0$ satisfying $t_0(r_{0}) = t({r})$. 
At this location, therefore, $\tau_i = 0$, meaning that the phase of the reconstructed echo is zero.
However, aberrations shift this center point away from the true position $r$ by an amount equal to the mean $\mu_{\text{A}_i}$ (Fig.~\ref{fig:Illustration}(a)). From the definition of $A_i$ in~\eqref{eq:Pulse} and~\eqref{eq:Image2}, together with the approximation in~\eqref{eq:TauApp}, we obtain
\begin{equation}
\mu_{\text{A}_i} = \frac{c_0}{2}\frac{\Delta t_{i}(r_{0})}{\cos\varphi_{\text{d},i}}, \quad
\sigma_{\text{A}_i} = \frac{c_0\sigma}{2\cos\varphi_{\text{d},i}}.
 \label{eq:MeanSDA} 
 \end{equation}
As we observe from these expressions, different acquisitions exhibit different displacements of the reconstructed echoes (Fig.~\ref{fig:Illustration}(b)), depending on the accumulated aberrations along the transmit and receive propagation paths and their angle difference $\varphi_\text{d}$. This difference in displacement is the reason why relative phase aberrations can be quantified from correlations. The reconstructed echoes are not merely shifted, but their width, defined by the standard deviation $\sigma_\text{A}$, also varies with the angle difference, independently of the aberrations (Fig.~\ref{fig:Illustration}(c)). This width determines the extent of the spatial-frequency band contributing to the image, whose center is given by ${k_c = 2\omega_c \cos\varphi_{\text{d}}/ c_0}$. Therefore, acquisitions sharing a common mid-angle probe overlapping but non-identical bands of spatial-frequency components
of tissue reflectivity, leading to a slight decorrelation that increases with larger differences between the two $\varphi_{\text{d},i}$. This effect will be analyzed in more detail in Section~\ref{sec:fluctuations}.

Using~\eqref{eq:MeanSDA}, we can compute the mean and standard deviation of $B$ analytically (see Appendix~\ref{appendix1}):
\begin{IEEEeqnarray}{rCl}
\mu_\text{B} & = & \frac{c_0}{2} \frac{\Delta t_{1}(r_{0})\cos\varphi_{\text{d},1} + \Delta t_{2}(r_{0})\cos\varphi_{\text{d},2}}{\cos^2\varphi_{\text{d},1} + \cos^2\varphi_{\text{d},2}},  \label{eq:MeanSDB}  \\
\sigma_\text{B}^2 & = & \frac{c^2_0}{4}\frac{\sigma^2}{\cos^2\varphi_{\text{d},1} + \cos^2\varphi_{\text{d},2}}.
\label{eq:SDB}
 \end{IEEEeqnarray}
To leverage the symmetry of $B$ for solving the integral in~\eqref{eq:Integrand1}, we define a new coordinate ${{\delta r_\text{B}} =  {r} - r_{0}  - \mu_\text{B}}$ centered around the mean of $B$ and replace~\eqref{eq:TauApp} and~\eqref{eq:MeanSDB} in~\eqref{eq:Integrand1} to obtain
\begin{equation}
{\langle C (r_0) \rangle} \!\approx \!\tilde{X} e^{i{\phi}(r_{0})}\!\!\int\! B({\delta r_\text{B}}) e^{-i\frac{2\omega_c}{c_0}(\cos\varphi_{\text{d},1} - \cos\varphi_{\text{d},2}){\delta r_\text{B}}} d{\delta r_\text{B}}.
 \label{eq:Integrand2}   
\end{equation}
The phase ${\phi}(r_{0})$ corresponds to the phase difference between reconstructed echoes at position $\mu_\text{B}$ (Fig.~\ref{fig:Illustration}(c)):
\begin{equation}
{\phi}(r_{0}) = \alpha\omega_c \left(\frac{\Delta t_{1}(r_{0})}{\cos\varphi_{\text{d},1}} - \frac{\Delta t_{2}(r_{0})}{\cos\varphi_{\text{d},2}}\right),
 \label{eq:IntegrandPhase}   
\end{equation}
with
\begin{equation} \alpha=\frac{\cos^2\varphi_{\text{d},1}\cos\varphi_{\text{d},2} + \cos\varphi_{\text{d},1}\cos^2\varphi_{\text{d},2}}{\cos^2\varphi_{\text{d},1} + \cos^2\varphi_{\text{d},2}}.
 \label{eq:IntegrandPhaseConstant}   
\end{equation}
When the image location $r_{0}$ is far from the probe relative to $\sigma_\text{B}$, the integration limits in~\eqref{eq:Integrand2}
can be approximated as $[-\infty, \infty]$. 
Appendix~\ref{appendix2} shows that, in this case, the integral in~\eqref{eq:Integrand2}
is real-valued, and the expected value of common mid-angle correlations reduces to  
\begin{equation}
\langle C({r_0}) \rangle \approx \tilde{X}\sqrt{2\pi}\sigma_\text{B} B_0(r_{0})  e^{i\phi({r_0})},
 \label{eq:Xcorr3}   
\end{equation}
where 
   \begin{equation}
 B_0(r_{0}) = \exp \left[\frac{-\phi(r_{0})^2}{2\omega^2_{c}\sigma^2 {\beta}}\right]
 \label{eq:B0_main}  
 \end{equation}
 is the peak value of the product of the two PSFs $A_i$, quantifying the local coherence between $I_1$ and $I_2$. The scalar factor $\beta$, defined in~\eqref{eq:beta}, depends on the angle differences $\varphi_{\text{d}, i}$ and accounts for the different widths of the PSFs. It reaches its maximum value $\beta=2$ when $\sigma_{\text{A}_1} = \sigma_{\text{A}_2}$ and decreases toward $1$ as their widths become increasingly different. 
 
The role of $B_0$ is discussed in detail in the next Section~\ref{sec:fluctuations}; here, we focus on the argument of the correlation, which provides direct access to the relative, local phase aberrations between the two images: 
\begin{equation}
\arg\langle C({r_0}) \rangle \approx \alpha \omega_c\left(\frac{\Delta t_{1}({r_0})}{\cos\varphi_{\text{d},1}} - \frac{\Delta t_{2}({r_0})}{\cos\varphi_{\text{d},2}}\right).
 \label{eq:XcorrPhaseFinal}   
\end{equation}
This phase is measured at the peak location of the product between the PSFs, which can be viewed as a normal probability distribution governing the phase measurement (Fig.~\ref{fig:Illustration}(c)). The term in parentheses arises from the relative shift of the reconstructed echo positions, whereas the factor $\alpha$ accounts for the stretching or shrinking of the spatial frequencies with the change of the transmit and receive angles.

\subsection{New Insights}
\label{sec:new_insights}

\begin{figure}[t]
\centering
\includegraphics[width=1\columnwidth]{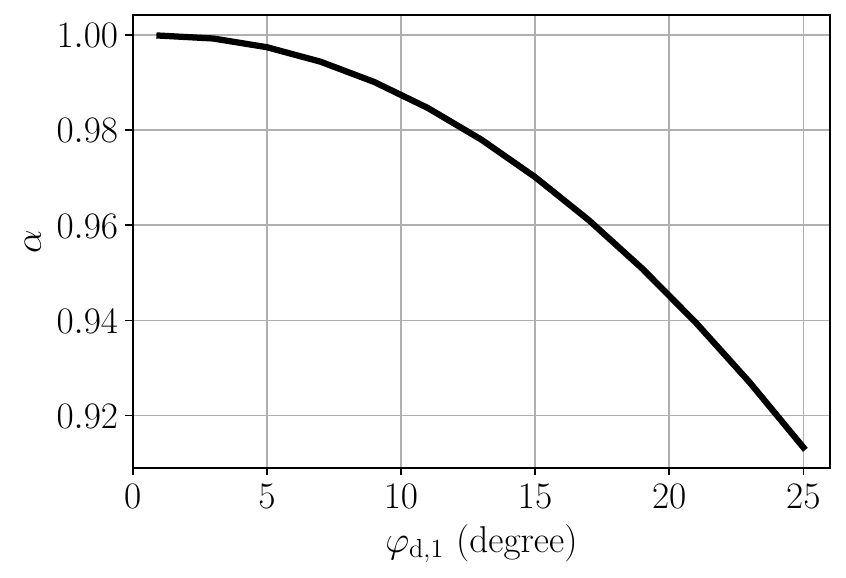}
\caption{Values of the factor $\alpha$ as a function of $\varphi_{\text{d},1}$, computed from~\eqref{eq:IntegrandPhaseConstant}, for   $\varphi_{\text{d},1} \in [1^\circ, 25^\circ]$ and ${\varphi_{\text{d},1} - \varphi_{\text{d},2} = 2^\circ}$. These angular parameters correspond to those used in~\cite{stahli2019forward}.}
\label{fig:AngularTerm}
\end{figure}

Using geometrical arguments, Staehli et al.~\cite{stahli2019forward} found a similar relationship between correlation phases and aberrations. Our result~\eqref{eq:XcorrPhaseFinal}, however, contains an angle-dependent factor $\alpha$ that was omitted in~\cite{stahli2019forward}, although the authors noted that the spatial frequencies vary with $\varphi_\text{d}$. While $\alpha\approx1$ is reasonable for small angles, larger values of $\varphi_\text{d}$ lead to increasing phase-modeling errors (Fig.~\ref{fig:AngularTerm}). In particular, $\alpha$ could play a significant role in the adaptation of CUTE to emerging multi-aperture systems with larger angular coverage~\cite{VeraThesis2025}.

\begin{figure}[t]
\centering
\includegraphics[width=1\columnwidth]{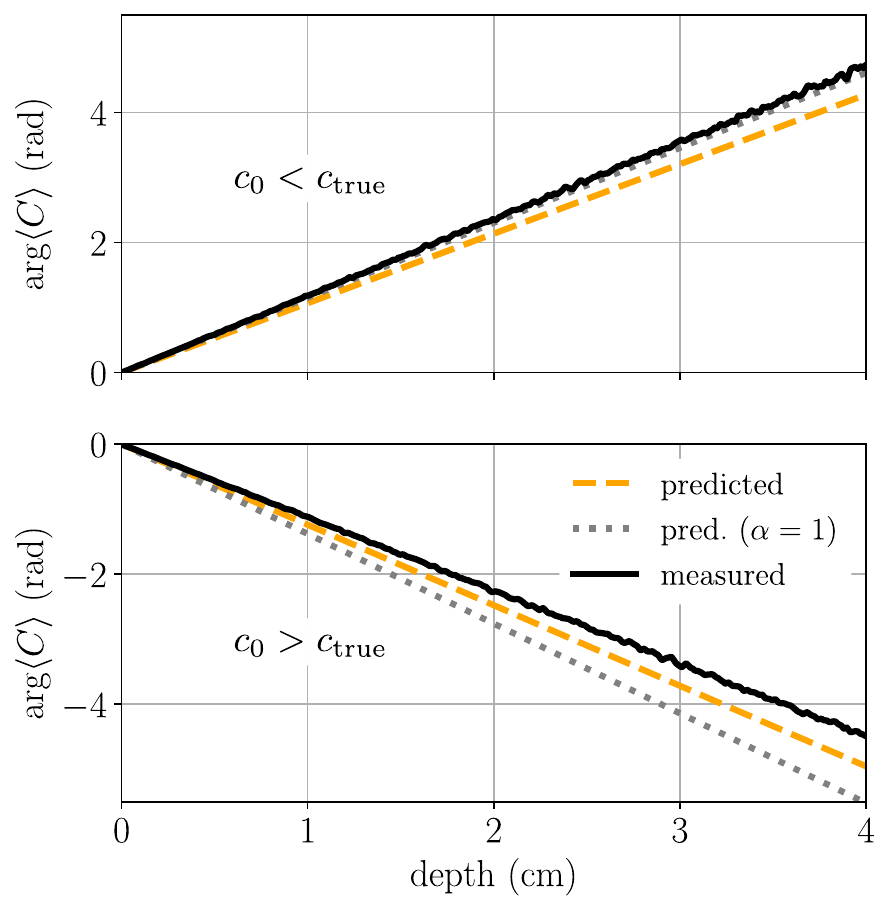}
\caption{Correlation phases measured from analytically computed images (black solid) versus the theoretical predictions from~\eqref{eq:XcorrPhaseFinal}, shown with (orange dashed) and without (gray dotted) the angle-dependent factor $\alpha$. The true medium has a uniform speed of sound ${c_\text{true} = 1525~\text{m/s}}$. The assumed speed of sound is $c_0 = 1400~\text{m/s}$ (top) and $c_0 = 1650~\text{m/s}$ (bottom), corresponding to the same absolute mismatch relative to $c_\text{true}$. These values span the typical range of speed of sound encountered in tissue. Images are computed using~\eqref{eq:Image2}, with travel times $t_0$ and $t$ obtained through~\eqref{eq:tt_law} using $c_0$ and $c_\text{true}$, respectively. The scattering function is modeled as randomly distributed point scatterers along the mid-angle direction, and the point-spread function $A$ follows~\eqref{eq:Pulse}, with a center frequency of $5~\text{MHz}$ and a fractional bandwidth of $50~\%$. 
Correlations are computed between images with $\varphi_\text{m} = 0^\circ$, $\varphi_{\text{d},1} = 25^\circ$, and $\varphi_{\text{d},2} = 23^\circ$ and averaged over $500$ realizations of diffuse scatterers. The acquisition parameters were chosen based on previous work in CUTE~\cite{stahli2019forward}. }
\label{fig:PhaseModel}
\end{figure}

Correlation phases are linked not to the aberrations at the true scatterer location~${r}$, but rather to those at the image location~$r_{0}$. This distinction is important for two reasons. First, the estimated local phase aberrations are inherently assigned to the erroneous echo positions, as previously noted by other authors~\cite{ali_sound_2023}. Second, the model~\eqref{eq:XcorrPhaseFinal} relies on the approximation~\eqref{eq:LinearApp}, which assumes small deviations $\delta {r}$ between ${r}$ and $r_{0}$, such that ${t(r_0 + \delta r) -  t(r_0)} \approx t_0(\delta {r})$. This approximation introduces a mismatch between the true and predicted deviations, related by $\delta {r} \propto (c_0/c_\text{true}) \delta {r}_\text{true}$. For strong aberrations, typically occurring at greater depths, the model~\eqref{eq:XcorrPhaseFinal} no longer describes the measured phases accurately. This effect, not previously addressed, is illustrated in Fig.~\ref{fig:PhaseModel}, which compares correlation phases measured from analytically computed images using~\eqref{eq:Image2} against the theoretical predictions from~\eqref{eq:XcorrPhaseFinal} in a homogeneous medium. For ${c_0<c_\text{true}}$, reconstructed echoes are systematically shallower than their true positions, and this deviation is underestimated by the approximation in~\eqref{eq:LinearApp}. As a result, modeled phases (orange dashed line) are smaller than the measured ones (black solid line) or, equivalently, shifted to deeper regions showing a reduced gradient versus depth. In contrast, for ${c_0>c_\text{true}}$, the misplacement of echoes is overestimated, yielding systematically larger absolute phase values at each location. Interestingly, Fig.~\ref{fig:PhaseModel} also shows that neglecting the angle-dependent factor $\alpha$ (gray dotted lines) makes phase modeling errors asymmetric with respect to $c_0$.

The approximation in~\eqref{eq:LinearApp} is essential to linearize the relationship between correlation phases~$\phi$ and phase aberrations~$\Delta t$. In addition, CUTE employs further linearizations by (i) treating~$\Delta t$ as the phase aberration at the true echo position and (ii) assuming straight rays to relate~$\Delta t$ to the tissue speed of sound~\cite{stahli2019forward}, neglecting refraction and finite-frequency effects~\cite{KortaMartiartu2020}. Even without the latter, echo mislocalization alone introduces nonlinearities in the relationship between~$\phi$ and the speed of sound. Therefore, future speed-of-sound imaging algorithms based on common-mid-angle correlations should use iterative strategies to progressively correct aberrations and reduce phase-modeling errors~\cite{Cui2026ADM, KortaMartiartu26}.

\subsection{Comment on Ensemble Averaging}
\label{sec:ensavg}
Equation~\eqref{eq:XcorrPhaseFinal} models the phase of the ensemble-averaged correlation $\left\langle C \right\rangle$. In practice, however, correlations are estimated in the vicinity of each image position $r_\text{c}$ from a single realization of disorder as
\begin{equation}
 \widehat{C}({r_\text{c}}) = \frac{1}{L }\int_{{r_\text{c}}-\frac{L}{2}}^{{r_\text{c}}+\frac{L}{2}}  I_1(r_{0})I^*_2(r_{0})  dr_{0}, 
 \label{eq:Xcorr2_practice}   
\end{equation}
with $L$ the size of the correlation kernel~\cite{stahli2019forward}.
This is justified if (i) tissue is spatially ergodic~\cite{Insana90},
and (ii) aberrations are constant over the correlation kernel $L$ (isoplanatism). Under these conditions, the correlation integral in~\eqref{eq:Xcorr2_practice} replaces the ensemble averaging in~\eqref{eq:Xcorr2}. Nevertheless, an inherent trade-off exists: a large $L$ improves convergence to the expected value $\left\langle C \right\rangle$ but may violate the assumption of isoplanatism, which generally holds only for small $L$ in heterogeneous tissues. This trade-off constrains the choice of $L$ and increases the uncertainty of phase estimates. The following section provides a detailed analysis of these uncertainties.

\section{Fluctuations of common-mid-angle correlation phases}
\label{sec:fluctuations}

Even in the absence of external noise and modeling errors, correlation-phase measurements fluctuate around their expected value~\eqref{eq:XcorrPhaseFinal} due to the intrinsic randomness of speckle. In this section, we quantify and discuss the fundamental accuracy limit imposed by speckle through a closed-form expression for the correlation-phase variance, derived in
Appendix~\ref{sec:derivation_variance}. We first present this result and discuss its physical interpretation (Section~\ref{sec:variance_interpretation}), before commenting on the additional contribution of clutter (Section~\ref{sec:clutter}).

\subsection{Correlation-Phase Variance and its Interpretation}
\label{sec:variance_interpretation}

With a slight abuse of notation, let $\phi$ denote the argument of the correlation ${\widehat{C}}$ estimated from a single realization using~\eqref{eq:Xcorr2_practice}, i.e.,
\begin{equation}
\phi({r_\text{c}}) = \arg\left( {\widehat{C}}({r_\text{c}})\right) = \arctan\left[\frac{\operatorname{Im} ({\widehat{C}}({r_\text{c}})) }{\operatorname{Re} ({\widehat{C}}({r_\text{c}}))}\right].
 \label{eq:XcorrPhaseAtan}   
\end{equation}
We use $\langle\phi\rangle$ to denote the phase of ensemble-averaged correlations $\langle C\rangle$ studied in Section~\ref{sec:theory}. To quantify the fluctuations of $\phi$ due to speckle, we compute its variance $\sigma^2_{\phi}$ by propagating the uncertainties of the real and imaginary parts of ${\widehat{C}}$: 
\begin{multline}
\sigma^2_{\phi} = \frac{1}{\left|\langle C \rangle \right|^2}\left[ \langle \operatorname{Im}({\widehat{C}})^2\rangle \cos^2\langle\phi\rangle + \langle \operatorname{Re}({\widehat{C}})^2\rangle \sin^2\langle\phi\rangle \right .\\\left. -  \langle \operatorname{Re}({\widehat{C}})\operatorname{Im}({\widehat{C}})\rangle \sin\left(2\langle\phi\rangle \right)\right].
 \label{eq:variancePhi}   
\end{multline}

Following similar derivations as in Section~\ref{sec:CMA_correlations}, Appendix~\ref{sec:derivation_variance} shows that the variance of correlation phases is explicitly given by 
\begin{equation}
\sigma^2_\phi({r_\text{c}}) \approx  
 \sqrt{\pi}\frac{\bar{\sigma}_\text{A}}{L}  \left[B^{-2}_0({r_\text{c}}) - 1\right],
 \label{eq:VarPhi_Final}    
\end{equation}  
where ${\bar{\sigma}_\text{A}}$ is the average width of the PSF along the mid-angle direction $\kvec_\text{m}$:
    \begin{equation}
\bar{\sigma}_\text{A} = \frac{c_0\sigma}{4}\sqrt{\frac{1}{\cos^2\varphi_{\text{d},1}} + \frac{1}{\cos^2\varphi_{\text{d},2}}}.
 \label{eq:avgPSFwidth} 
 \end{equation}    

\begin{figure}[t]
\centering
\includegraphics[width=1\columnwidth]{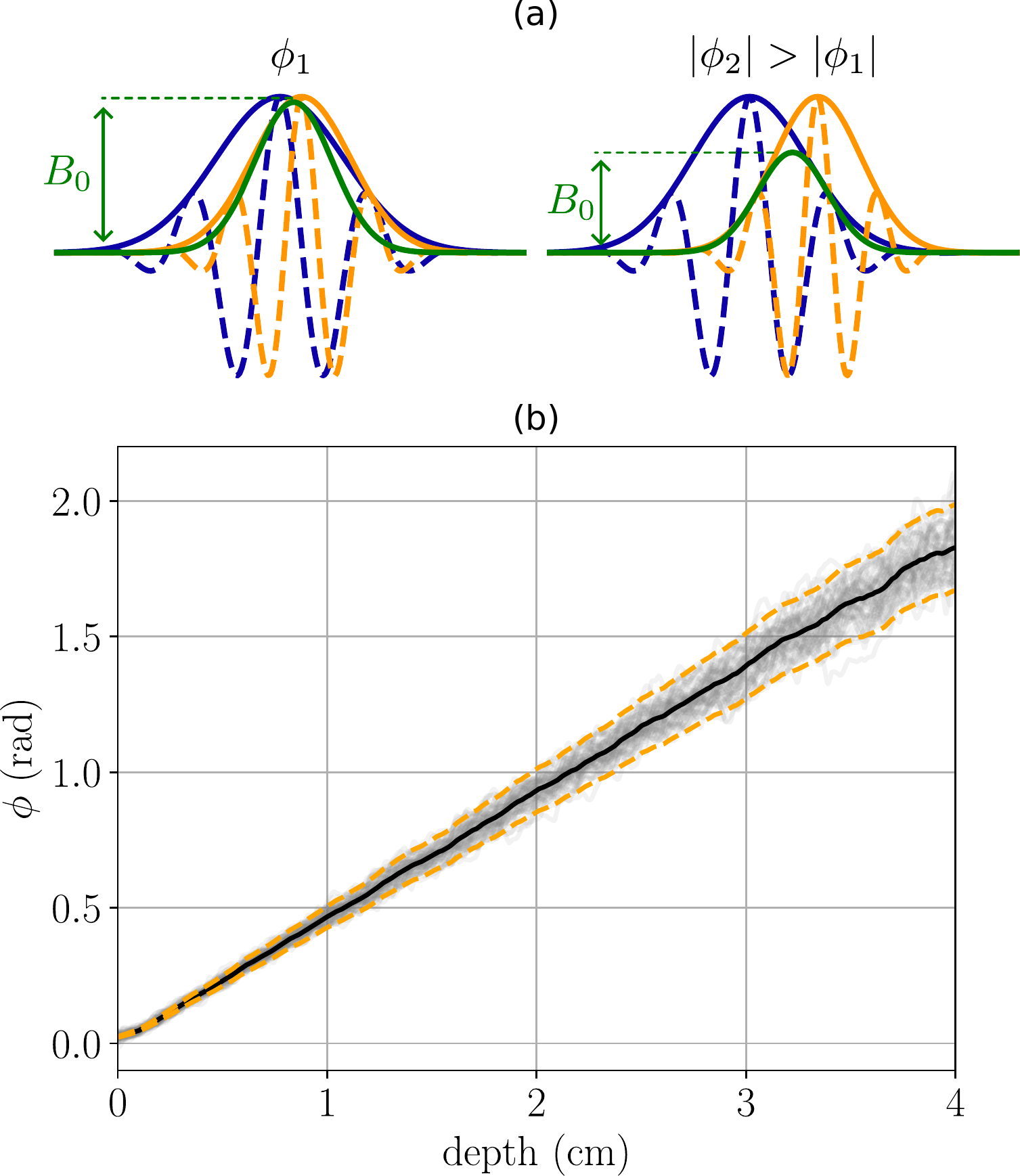}
\caption{Influence of phase aberrations on the variance of correlation phases $\phi$. (a) Illustration of the coherence measure $B_0$ and its decay when $\phi$ increases in magnitude. (b) Observed (gray) and predicted (orange) variability of correlation phases for analytically computed images as in Fig.~\ref{fig:PhaseModel}. We use the same acquisition parameters, but with an assumed speed of sound of $c_0=1475~\text{m/s}$ to avoid phase wrapping. Correlations are estimated using $\widehat{C}$ defined in~\eqref{eq:Xcorr2_practice}, with a correlation-kernel size $L=2$~mm. The black line is the mean phase across $50$ realizations, whereas gray lines show individual realizations. Orange lines indicate the 95\% confidence interval predicted from~\eqref{eq:VarPhi_Final}. As aberrations increase with depth, the correlation-phase variance increases accordingly.}
\label{fig:VarianceModel}
\end{figure}

\begin{figure*}[t!]
\centering
\includegraphics[width=0.7\textwidth]{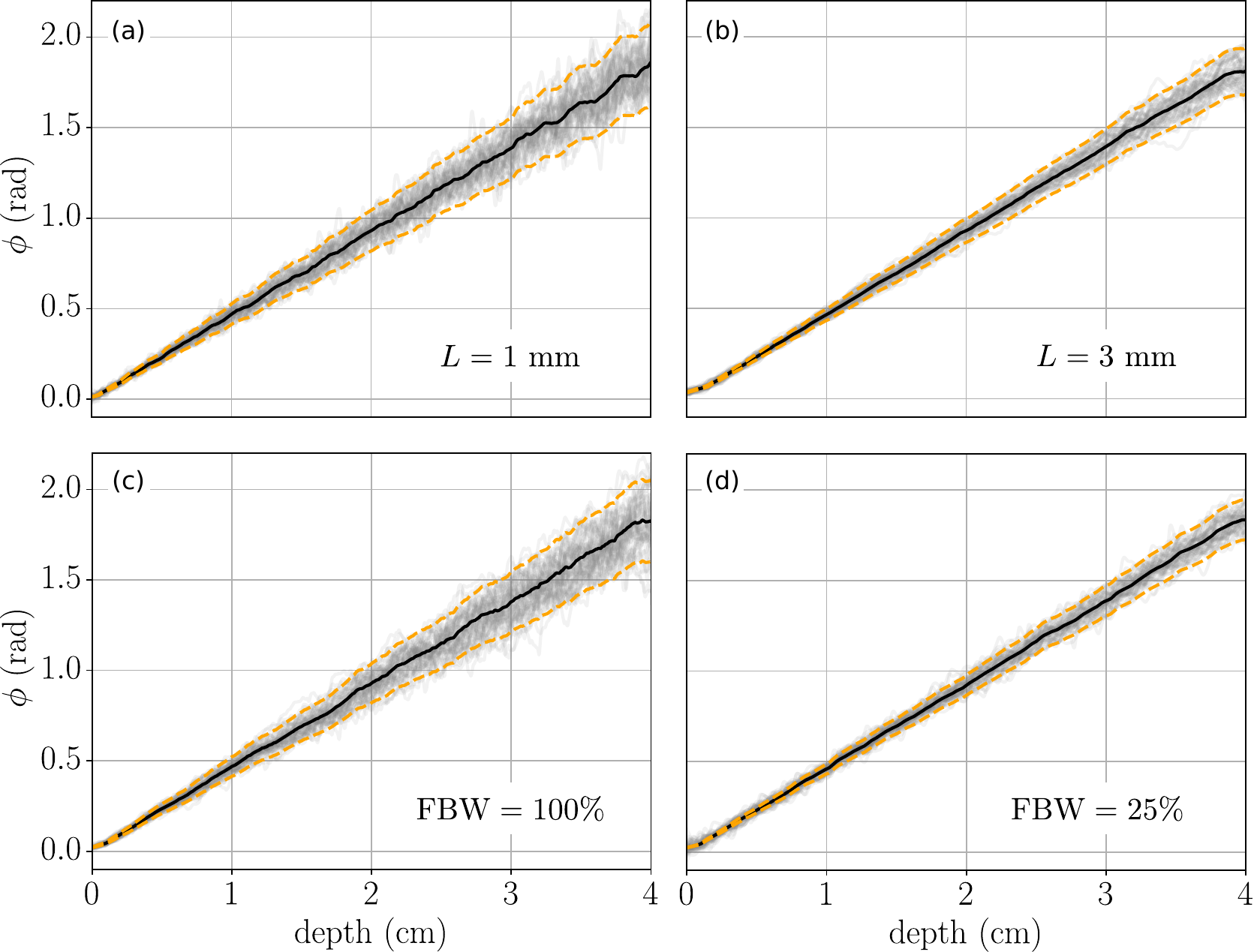}
\caption{Influence of the correlation-kernel size $L$ and fractional bandwidth (FBW) on the variance of correlation phases $\phi$. Data and reconstruction parameters from~\ref{fig:VarianceModel}(b) are used as reference. (a)-(b) Correlation phases are computed with a kernel size of $L=1~\text{mm}$ and  $L=3~\text{mm}$, respectively. (c)-(d) Data is simulated with a fractional bandwidth of $100~\%$ and $25~\%$, respectively. In each panel, the black line represents the mean phase over $50$ realizations, gray lines show phase estimates for each realization, and orange lines indicate the 95\% confidence interval predicted from~\eqref{eq:VarPhi_Final}. }
\label{fig:VarianceChange}
\end{figure*}

Several conclusions can be drawn from the expression of $\sigma^2_\phi$ in~\eqref{eq:VarPhi_Final}.
First, the variance of correlation-phase measurements scales as $1/N$, where $N = L/\bar{\sigma}_\text{A}$ is the number of resolution cells within the correlation kernel. This is consistent with Section~\ref{sec:ensavg}: if tissue is spatially ergodic, each resolution cell acts as an independent realization of disorder, so the correlation integral approximates the ensemble average. As $N$ increases, correlation phases more closely match their expected value.

For a fixed $L$, the number of resolution cells increases when $\bar{\sigma}_\text{A}$ decreases.   
The PSF becomes narrower when the emitted signals have a larger bandwidth ($\propto 1/\sigma$) and data is acquired with a smaller angle difference $\varphi_{\text{d},i}$. As a result, phase estimates are noisier for larger $\varphi_{\text{d},i}$, an effect observed in~\cite{Salemi23} (see their Fig. 2).

While the dependence on $N$ is largely intuitive and has been discussed previously~\cite{lambert_reflection_2020, stahli2019forward}, our result in~\eqref{eq:VarPhi_Final} reveals that aberrations directly affect the correlation-phase variance. Aberrations appear through the term $B^2_0$, which measures the intensity at the location where correlation phases are extracted (Fig.~\ref{fig:VarianceModel}(a)). Essentially, $B^2_0$ quantifies the degree of coherence between the reconstructed images, and the phase variance scales with the relative loss of coherence, expressed as $B^{-2}_0 - 1 = (1-B^2_0)/B^2_0$.  As shown in~\eqref{eq:B0_main}, $B^2_0$ decreases with $|\phi|$; thus, the phase variance grows with the magnitude of the phases themselves. The stronger the aberrations, the larger the shift between reconstructed echoes and the lower the peak intensity $B^2_0$ (Fig.~\ref{fig:VarianceModel}(a)), ultimately leading to a larger uncertainty in phase estimation (Fig.~\ref{fig:VarianceModel}(b)). If the magnitude of unwrapped phases is small relative to ${\sigma\omega_c}$ (i.e., the inverse of the fractional bandwidth),
the variance becomes linearly related to $\langle\phi\rangle^2$:
\begin{equation}
\sigma^2_\phi({r_\text{c}}) \approx  
  \frac{\sqrt{\pi}}{N\omega_c^2\sigma^2{\beta}}\langle\phi({r_\text{c}})\rangle^2,
 \label{eq:SDPhi_linear}    
\end{equation}
This result implies that, in the absence of aberrations (i.e., $\langle\phi\rangle=0$),  uncertainties in common-mid-angle correlation phases become negligible. This is consistent with the same effect observed in the phase shifts between two arbitrary speckle signals~\cite{goodman_speckle_2007}. Note that the correlation-phase variance is a local quantity; it becomes negligible whenever the local correlation phase is zero.

Strikingly,~\eqref{eq:SDPhi_linear} shows that phase uncertainty increases with the fractional bandwidth. This may appear counterintuitive, as a larger bandwidth narrows the PSFs and increases $N$, as discussed above. However, for a fixed $\phi$, a broader bandwidth also reduces the overlap between the two PSFs, thereby decreasing $B^2_0$. Therefore, aberrations cause coherence to decay more rapidly with increasing bandwidth. The effect of both bandwidth and kernel size on $\sigma^2_\phi$ is shown in Fig.~\ref{fig:VarianceChange}. Although reducing the bandwidth as much as possible may seem attractive for minimizing $\sigma^2_\phi$, the model in~\eqref{eq:VarPhi_Final} no longer holds when $L$ becomes comparable to the PSFs (see~\eqref{eq:EnsIm2_2sol}).

\subsection{Comment on the Influence of Clutter}
\label{sec:clutter}

Equation~\eqref{eq:VarPhi_Final} describes the aberration-induced phase variance and defines the achievable phase accuracy in the absence of noise. We have shown that this variance is proportional to the loss of coherence caused by phase aberrations, although the same reasoning extends to any other factor that degrades coherence, including clutter arising from thermal and electronic noise, multiple scattering, or reverberations. A detailed analysis of clutter-induced variance is beyond the scope of this work; nevertheless, two remarks can be made regarding its expected behavior. First, clutter is often largely uncorrelated across different transmit and receive plane waves. In practice, slight transmit and receive focusing is typically applied through coherent compounding before computing the correlations~\cite{stahli2019forward}. This focusing reduces clutter and thus its contribution to the phase variance. Second, part of the clutter is expected to be spatially uncorrelated, so its variance contribution scales inversely with the number of resolution cells $N$. As a result, increasing the bandwidth may decrease the total variance, potentially outweighing the increase predicted by~\eqref{eq:SDPhi_linear}. The following section experimentally validates the predicted variance and further illustrates the joint effect of aberrations and clutter.

%% file: inputs/Experiment_clean.tex
\section{Experimental validation}
\label{sec:experiment}

In this section, we aim to evaluate how well the model in~\eqref{eq:VarPhi_Final} explains the fluctuations observed in experimental correlation-phase measurements and whether the predicted dependence of the variance on the correlation phase, kernel size, and bandwidth is observed experimentally.

\begin{figure*}[t!]
\centering
\includegraphics[width=1\textwidth]{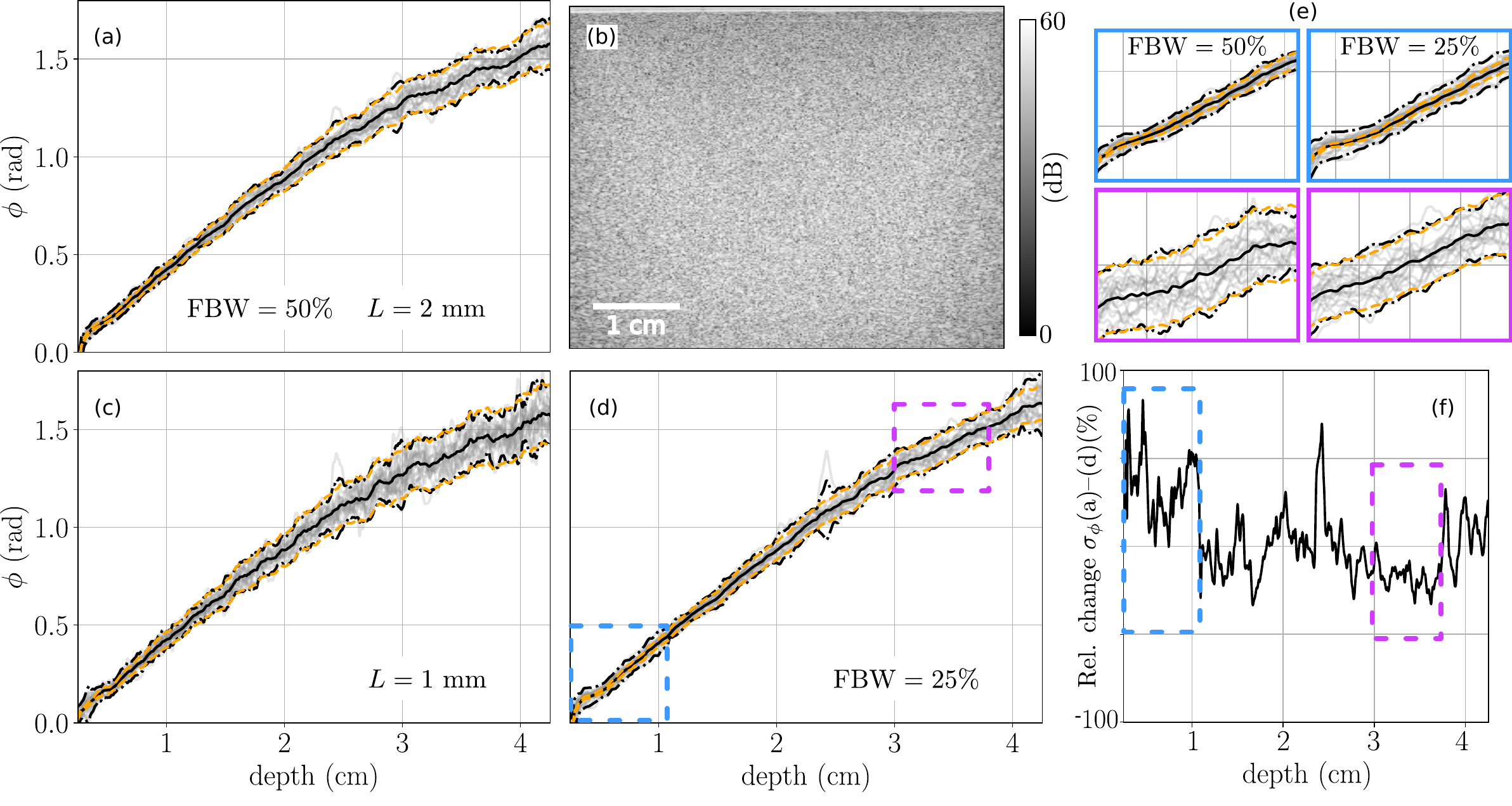}
\caption{Experimentally measured (black dashed) and theoretically predicted (orange dashed) variance of correlation phases $\phi$. The medium, shown in (b), is a uniform tissue-mimicking phantom with a speed of sound of 1518~m/s. The assumed speed of sound is 1400~m/s. (a)~Observed fluctuations of correlation phases with depth for 20 realizations of disorder (gray lines); the mean phase is shown as a black solid line. Correlations are computed using $\varphi_\text{m} = 0$, $\varphi_{\text{d},1} = 9^\circ$, $\varphi_{\text{d},2} = 7^\circ$, a kernel size $L=2~\text{mm}$, and the fractional bandwidth (FBW) of the dataset (50\%). (c), (d) Same as in (a), but with half the kernel size and half the FBW, respectively. (e) Zoomed views of (a) and (d) at the locations marked by dashed boxes in (d). (f) Relative change in the observed phase fluctuations when reducing the FBW from $50\%$ (a) to $25\%$ (d), computed as ${(\sigma_\phi^{(d)}-\sigma_\phi^{(a)})/\sigma_\phi^{(a)} \! \times \! 100}$. Smaller $\varphi_\text{d}$ values than in simulations are used to ensure illumination at larger depths while avoiding edge waves. }
\label{fig:ExperimentalVariance}
\end{figure*}

We use the same dataset previously employed for calibration in~\cite{Korta_Martiartu_2024}. Data were acquired in a tissue-mimicking phantom (CIRS Inc. Norfolk, VA, USA) with uniform scattering properties and constant speed of sound of 1518~m/s (Fig.~\ref{fig:ExperimentalVariance}(b)). The experimental setup consisted of a SuperSonic\textsuperscript{\tiny\textregistered} MACH\textsuperscript{\tiny\textregistered} 30 ultrasound system (Hologic\textsuperscript{\tiny\textregistered} - Supersonic Imagine\textsuperscript{\tiny\textregistered}, Aix-en-Provence, France) equipped with an L18-5 linear probe containing 256 elements with a pitch of 0.2 mm. Steered plane-wave emissions were used at angles ranging from -27.5$^\circ$ to 27.5$^\circ$, with an angular step of 0.5$^\circ$, and the center frequency was 6.5 MHz with 50\% fractional bandwidth. This dataset contains 20 repeated measurements at different locations of the phantom to mimic different realizations of disorder.

For our analysis, we reconstruct transmit-receive plane-wave images according to~\eqref{eq:Image}--\eqref{eq:kmid} using the same set of angles as in transmission. As discussed in Section~\ref {sec:clutter}, we apply coherent compounding to reduce clutter before computing correlations. Following CUTE~\cite{stahli2019forward}, we define compounding angles ranging from $-25^\circ$ to $25^\circ$ in steps of $2^\circ$ and synthetically focus in emission and reception using a Gaussian weighting function (standard deviation: $3^\circ/\sqrt{2}$). Due to the lateral resolution achieved through compounding, we restrict our analysis to the mid-lateral position. Finally, correlations are computed using~\eqref{eq:Xcorr2} at the mid-angle $\varphi_\text{m} = 0$ to illustrate how the correlation phases evolve with depth, as aberrations increase. 

As shown in Fig.~\ref{fig:ExperimentalVariance}(a), the theoretical predictions based on the mean phase values closely match the experimentally measured phase variance. This result confirms that phase estimation becomes more uncertain when aberrations increase in magnitude, a key finding of this work. Deviations from the theoretical model mainly occur when the correlation phases are relatively small, approximately within the first centimeter of depth. Here, the measured $\sigma_\phi$ is on average about 150\% higher than the theoretical prediction, whereas at larger depths the relative difference drops below 1\%. Near the probe, the aberration-induced phase variance is small, and clutter likely becomes the dominant contribution to the total phase variance.

Because both aberration- and clutter-induced variances scale inversely with the number of resolution cells, the relative difference between measured and predicted $\sigma_\phi$ remains unchanged when the kernel size is reduced by half (Fig.~\ref{fig:ExperimentalVariance}(c)). However, the variance reduction for a narrower bandwidth observed in Fig.~\ref{fig:VarianceChange} is less evident in Fig.~\ref{fig:ExperimentalVariance}(d). Compared to the full-bandwidth case (Fig.~\ref{fig:ExperimentalVariance}(e)), $\sigma_\phi$ appears to decrease slightly, e.g., by about 20\% on average at depths of 3-4 cm, but it becomes nearly 30\% larger at shallow regions (Fig.~\ref{fig:ExperimentalVariance}(f)). This clearly reflects the two competing mechanisms discussed earlier. While narrowing the bandwidth reduces aberration-induced variance, it also decreases the number of resolution cells, making the influence of clutter on phase estimates more pronounced. 

These experimental results confirm that the theoretical model developed here provides a lower bound on the correlation-phase variance and therefore tells us how accurately correlation phases can, in principle, be estimated. The model could be used to construct a data-covariance operator for speed-of-sound inversion that weights phase measurements according to their reliability in order to improve robustness.

%% file: inputs/Conclusion_clean.tex
\section{Conclusion}
\label{sec:conclusion}

We have developed a theoretical formalism to relate common-mid-angle speckle correlations to the phase aberrations caused by speed-of-sound inhomogeneities. This formalism (i) completes and clarifies previously established models describing a linear relationship between correlation phases and the mismatches between the assumed and true travel times and (ii) enables the prediction of correlation-phase fluctuations resulting from aberration-induced speckle decorrelation. 

For image pairs obtained with transmit–receive angles sharing a common mid-angle, phase aberrations manifest as displacements of reconstructed echoes along the same direction. The relative shift in echoes between the two images is what determines the correlation values at each location. We have shown that the phase of the correlations scales linearly with the relative echo shift, whereas the correlation magnitude, which reflects the coherence between the two images, decays exponentially with the square of correlation phases. A second-order statistical analysis revealed that this loss of coherence with increasing phase magnitude reduces the reliability of correlation-phase estimates, a phenomenon we have experimentally verified in a uniform tissue-mimicking phantom. 

The variance of the phase difference between two speckle signals is fundamentally governed by their coherence~\cite{goodman_speckle_2007}, while averaging over $N$ resolution cells further reduces this variance as $\sim 1/N$. It is therefore natural that Eq.~\eqref{eq:VarPhi_Final} shares the same general structure as variance expressions derived for other aberration-estimation methods. Bureau~et~al.~\cite{bureau_three-dimensional_2023}, for instance, obtained ${\sigma_\phi^2 \propto 1/(\mathcal{C}^2N)}$ for aberration laws estimated through the distortion matrix approach, where $\mathcal{C}$ is a coherence factor. What differs across methods is the definition of coherence itself, which depends on the correlations used to estimate the aberrations. For common-mid-angle correlations, the corresponding coherence factor $B_0$ can be expressed explicitly in terms of the local correlation phase $\phi$, and therefore of the phase aberrations accumulated by the transmit and receive waves. Since these aberrations are precisely the observables used to retrieve the speed-of-sound distribution, our result provides an uncertainty model that could be incorporated into the inversion through a Bayesian formulation~\cite{Tarantola2005}.
The formalism developed here thus provides both physical insight and a theoretical basis for advancing speed-of-sound imaging. 
More broadly, it establishes a benchmark for comparing the fundamental accuracy limits of more advanced aberration-estimation techniques, including those based on time-reversal analysis~\cite{Lambert2020, Bendjador2020} and the Radon transform~\cite{Beuret24}.

%% file: inputs/Appendix_clean.tex
\appendices

\section{Product of two Gaussian functions}
\label{appendix1}
We define the product between two Gaussian functions as
\begin{equation*}
B(x, y) = A_1(x) A_2(x - y) \quad \text{with} \quad A_i(x) = e^{-(x - \mu_i)^2/2\sigma^2_i}.
 \label{eq:Gaussi}  
 \end{equation*}
It can be shown that $B$ is also a Gaussian function:
\begin{equation}
B(x, y) =   B_0(y) e^{-(x - \mu_\text{B})^2/2\sigma_\text{B}^2},
 \label{eq:EnvB}   
\end{equation}
where its mean and standard deviation are
\begin{equation}
\mu_{\text{B}} = \frac{\mu_1\sigma_2^2 + (\mu_2 + y)\sigma^2_1}{\sigma^2_1 + \sigma^2_2}; \quad \sigma_{\text{B}}^2 = \frac{\sigma^2_1 \sigma^2_2}{\sigma^2_1 + \sigma^2_2}  
 \end{equation}
and its amplitude is
\begin{equation}
B_0(y) = e^{-( y - (\mu_1 - \mu_2))^2/2(\sigma^2_1 + \sigma^2_2)}.
 \end{equation}

\section{Solution of the integral in~\eqref{eq:Integrand2}}
\label{appendix2}
In~\eqref{eq:Integrand2}, we have an integral of the form
\begin{equation}
\int_{-\infty}^{\infty} B(u) e^{ibu} du  \quad \text{with} \quad B(u) = B_0 e^{-u^2/2 \sigma^2_\text{B}}
 \label{eq:AppIntegral}  
 \end{equation}
and $b$ a constant value. Since the integrand can be written as
\begin{equation}
B_0 e^{-b^2 \sigma^2_\text{B}/2} e^{-(u - ib\sigma^2_\text{B})^2/2 \sigma^2_\text{B}},
 \label{eq:AppIntegrand}  
 \end{equation}
 the solution of~\eqref{eq:AppIntegral} is
 \begin{equation}
\int_{-\infty}^{\infty} B(u) e^{ibu} du = \sqrt{2\pi}\sigma_\text{B}B_0e^{-b^2 \sigma^2_\text{B}/2},
 \label{eq:AppIntegrandSol}  
 \end{equation}
which is real-valued. 

Specifically, the solution of the integral in~\eqref{eq:Integrand2} is
\begin{equation}
\int B({\delta r_\text{B}}) e^{-i\frac{2\omega_c}{c_0}(\cos\varphi_{\text{d},1} - \cos\varphi_{\text{d},2}){\delta r_\text{B}}} d{\delta r_\text{B}} \approx \sqrt{2\pi}\sigma_\text{B} B_0(r_{0})
 \label{eq:AppIntegralSol1} 
\end{equation}
 with $\sigma_\text{B}$ given by~\eqref{eq:SDB},
   \begin{equation}
 B_0(r_{0}) = \exp \left[\frac{-\phi(r_{0})^2}{2\omega^2_{c}\sigma^2 {\beta}}\right],
 \label{eq:AppIntegralSol2}  
 \end{equation}
 and
\begin{equation}
{\beta} = \frac{(\cos\varphi_{\text{d},1} {+} \cos\varphi_{\text{d},2})^2}{\cos\varphi^2_{\text{d},1} + \cos\varphi^2_{\text{d},2}}.
\label{eq:beta}
 \end{equation}
The term $\phi(r_{0})$ is the correlation phase defined in~\eqref{eq:IntegrandPhase}, and~\eqref{eq:AppIntegralSol1} assumes small angular differences between correlated images, as is common practice to avoid phase wrapping when $|\phi|\geq\pi$.

\section{Analytical Derivation of Correlation-Phase Variance}
\label{sec:derivation_variance}

Equation~\eqref{eq:variancePhi} shows that the correlation-phase variance $\sigma^2_{\phi}$ is composed of three terms. Since they
  require similar calculations, we derive $\langle \operatorname{Im}({\widehat{C}})^2\rangle$ here and provide only the results for $\langle \operatorname{Re}({\widehat{C}})^2\rangle$ and $\langle \operatorname{Re}({\widehat{C}})\operatorname{Im}({\widehat{C}})\rangle$. All these terms involve fourth-order correlations of the scattering function $\tilde{\chi}$, which is modeled as a zero-mean real Gaussian random field~\cite{Wagner88, Insana90}:
\begin{multline}
\langle \tilde{\chi}({r}) \tilde{\chi}({r^{\prime}}) \tilde{\chi}({r^{\prime\prime}}) \tilde{\chi}({r^{\prime\prime\prime}}) \rangle  = \tilde{X}^2 [\delta({r - r^{\prime}})\delta({r^{\prime\prime} - r^{\prime\prime\prime}}) \\+ \delta({r - r^{\prime\prime}})\delta({r^{\prime} - r^{\prime\prime\prime}}) + \delta({r - r^{\prime\prime\prime}})\delta({r^{\prime} - r^{\prime\prime}})].
 \label{eq:4rthGaussian}    
\end{multline}

 As a result, $\langle \operatorname{Im}({\widehat{C}})^2\rangle$ is composed of three terms:
\begin{multline}
 \langle \operatorname{Im}({\widehat{C}})^2\rangle  =   \frac{1}{L^2}\iint dr_{0} dr^\prime_{0} \underbrace{\langle \operatorname{Im}({I}_{1} {I}_{2}^*)\operatorname{Im}({I^{\prime}}_{1} {I^{\prime*}}_{2}) \rangle}_{\mathcal{P}_1 + \mathcal{P}_2 + \mathcal{P}_3}.
 \label{eq:EnsIm2}    
\end{multline}
For brevity, we denote $I_i := I_i(r_0)$ and $I^\prime_i := I_i(r^\prime_0)$ and use the same convention for $A_i$ and $\tau_i$ in what follows.

Using the expression for the image $I(r_{0})$ in~\eqref{eq:Image2}, the first term $\mathcal{P}_1$ is given by
\begin{multline}
\mathcal{P}_1{(r_0, r^\prime_0)} = \tilde{X}^2 \! \int \! d{r} A_1({r})A_2({r})  \sin[\omega_c(\tau_1({r}) - \tau_2({r}))]\\ \times \int d{r^\prime} A_1^{\prime}({r^\prime})A_2^{\prime}({r^\prime})\sin[\omega_c(\tau_1^{\prime}({r^\prime}) - \tau_2^{\prime}({r^\prime}))].
 \label{eq:EnsIm2_1}    
\end{multline}
Each integral is equal to the imaginary part of~\eqref{eq:Integrand1}. Thus, we can use the expression of $\langle C \rangle$ found in~\eqref{eq:Xcorr3} to obtain:
\begin{equation}
\frac{1}{L^2}\iint dr_{0} dr^\prime_{0} \mathcal{P}_1{(r_0, r^\prime_0)}  = \left|\langle C \rangle \right|^2 \sin^2\langle\phi ({r_\text{c}})\rangle, 
 \label{eq:EnsIm2_1sol}    
\end{equation}
where 
\begin{equation}
\left|\langle C \rangle \right|^2 =  2\pi\sigma^2_\text{B} B^2_0(r_{0})\tilde{X}^2,
\end{equation}
with $B_0$ defined in~\eqref{eq:AppIntegralSol2}. 

In contrast to the first term, the variance contributions of $\mathcal{P}_2$ and $\mathcal{P}_3$ arise from the interference between reconstructed echoes at different locations. This interference depends on the relative travel-time differences $\tau_i$ between the locations $r_{0}$ and $r^\prime_{0}$. It is therefore convenient to write $\tau_i^{\prime}$ in terms of $\tau_i$ as
\begin{equation}
\tau^{\prime}_i = \tau_i + \frac{2}{c_0}\cos\varphi_{\text{d},i}(r^\prime_{0} - r_{0}),
 \label{eq:TauAppprimer} 
 \end{equation} 
 where we again apply the approximation in~\eqref{eq:LinearApp}. 
 
 The terms $\mathcal{P}_2$ and $\mathcal{P}_3$ also involve products of PSFs of the form $A_iA_j^{\prime}$. Using~\eqref{eq:TauAppprimer}, the definition of $\tau_i$ in~\eqref{eq:TauApp}, and Appendix~\ref{appendix1}, this product reduces to the Gaussian function
 \begin{multline}
A_iA_j^{\prime} = \exp\left\{-\frac{((r^\prime_{0}-r_{0}) - (\mu_{\text{A}_i} - \mu_{\text{A}_j})) ^2}{8\bar{\sigma}_\text{A}^2}\right\} \\\times \exp\left\{-\frac{(\delta {r} - \mu_{\mathcal{T}})^2}{2\sigma^2_{\text{B}}}\right\},
\label{eq:AmpTermIII}
\end{multline}
with mean 
\begin{equation}
\mu_{\mathcal{T}} = \mu_{\text{B}} + \frac{(r^\prime_{0}-r_{0})\sigma^2_{\text{A}_i}}{4\bar{\sigma}_\text{A}^2}
\end{equation}
and effective average width ${\bar{\sigma}_\text{A} := \sqrt{\sigma^2_{\text{A}_i} + \sigma^2_{\text{A}_j}}/2}$. The means $\mu_{\text{A}_i}$, $\mu_{\text{B}}$ and the standard deviations $\sigma_{\text{A}_i}$, $\sigma_{\text{B}}$ are previously defined in~\eqref{eq:MeanSDA}--\eqref{eq:SDB}. 

We focus now on the derivation of the second term, i.e.,
\begin{multline}
\mathcal{P}_2{(r_0, r^\prime_0)} = \tilde{X}^2 \iint d{r} d{r^\prime} A_1({r})A_1^{\prime}({r})A_2({r^\prime}) A_2^{\prime}({r^\prime}) \\ \times \sin[\omega_c(\tau_1({r}) - \tau_2({r^\prime}))] \sin[\omega_c(\tau^\prime_1({r}) - \tau_2^\prime({r^\prime}))]. 
 \label{eq:EnsIm2_2}    
\end{multline}
Although the two integration variables ${r}$ and ${r^\prime}$ appear mixed within the sine functions, they can be separated using trigonometric identities. In addition, we express $\tau_i^{\prime}$ via~\eqref{eq:TauAppprimer} and perform the variable changes ${\delta r^{(\prime)}_{\mathcal{T}}} = r^{(\prime)} - r_{0} - \mu_{\mathcal{T}}$ to transform the sine product into
\begin{multline}
\frac{1}{2}\left\{ \cos\left[\frac{2\omega_c}{c_0}(\cos\varphi_{\text{d},1} - \cos\varphi_{\text{d},2})(r^\prime_{0} - r_{0})\right]\right. \\ - \left.  \cos\left[\frac{4\omega_c}{c_0}\cos\varphi_{\text{d},1}{\delta r^{}_{\mathcal{T}}}\right]\cos\left[\frac{4\omega_c}{c_0}\cos\varphi_{\text{d},2}{\delta r^{\prime}_{\mathcal{T}}}\right] \right. \\ -  \left.  \sin\left[\frac{4\omega_c}{c_0}\cos\varphi_{\text{d},1}{\delta r^{}_{\mathcal{T}}}\right]\sin\left[\frac{4\omega_c}{c_0}\cos\varphi_{\text{d},2}{\delta r^{\prime}_{\mathcal{T}}} \right] \right\}. 
\label{eq:sinesep2}
\end{multline}
The first term approximates to 1 when $\varphi_1$ and $\varphi_2$ are very close, as assumed here. Unlike the other terms, which oscillate rapidly with the integration variable, this first term is constant with respect to ${\delta r^{}_{\mathcal{T}}}$ and ${\delta r_{\mathcal{T}}^{\prime}}$; thus, it dominates the integral in~\eqref {eq:EnsIm2_2}. 
 With this approximation, substituting~\eqref{eq:AmpTermIII} into~\eqref{eq:EnsIm2_2} yields
\begin{equation}
\mathcal{P}_2{(r_0, r^\prime_0)}  \approx \frac{\tilde{X}^2}{2} \pi\sigma_{\text{A}_1}\sigma_{\text{A}_2} \exp\left\{-\frac{(r^\prime_{0}-r_{0})^2} {4\sigma_\text{B}^2}\right\}. 
 \label{eq:EnsIm2_5}    
\end{equation}
Provided that the size of the correlation kernel $L$ is significantly larger than $\sigma_\text{B}$ (e.g., $L \approx 32\sigma_\text{B}$ in the example used in Fig.~\ref{fig:PhaseModel}), we finally obtain 
\begin{equation}
 \frac{1}{L^2}\iint dr dr' \mathcal{P}_2{(r_0, r^\prime_0)} \approx \frac{ \left|\langle C \rangle \right|^2  }{LB^2_0({r_\text{c}})} \sqrt{\pi} \bar{\sigma}_\text{A},
 \label{eq:EnsIm2_2sol}    
\end{equation}
where we used~\eqref{eq:Xcorr3} to write the solution in terms of $\left|\langle C \rangle \right|^2$.

The exact same steps and arguments can be used to derive $\mathcal{P}_3$ and finally obtain
\begin{multline}
\frac{\langle \operatorname{Im}({\widehat{C}})^2\rangle}{\left|\langle C \rangle \right|^2} \approx  \sin^2\langle\phi\rangle  + \sqrt{\pi}\frac{ \bar{\sigma}_\text{A}}{L}  \left(B_0^{-2}  -  \cos\left(2\langle\phi\rangle\right)\right) .
 \label{eq:EnsIm2_Final}    
\end{multline}
Note that all the quantities are evaluated at the measurement location ${r_\text{c}}$, which was omitted to simplify the notation.

In the same manner, we find the closed forms of $\langle \operatorname{Re}({\widehat{C}})^2\rangle$ and $\langle \operatorname{Re}({\widehat{C}})\operatorname{Im}({\widehat{C}})\rangle$:
\begin{multline}
\frac{\langle \operatorname{Re}({\widehat{C}})^2\rangle}{\left|\langle C \rangle \right|^2} \approx  \cos^2\langle\phi\rangle +  \sqrt{\pi}\frac{ \bar{\sigma}_\text{A}}{L}  \left(B_0^{-2}   +  \cos\left(2\langle\phi\rangle\right)\right)
 \label{eq:EnsRe2_Final}    
\end{multline}
and
\begin{equation}
\frac{\langle \operatorname{Re}({\widehat{C}})\operatorname{Im}({\widehat{C}})\rangle}{\left|\langle C \rangle \right|^2} \approx   \sin\left(2\langle\phi\rangle\right)  \left( \frac{1}{2} +  \frac{1}{L}\sqrt{\pi}\bar{\sigma}_\text{A}\right).
 \label{eq:EnsImRe_Final}    
\end{equation}

Finally, substituting~\eqref{eq:EnsIm2_Final},~\eqref{eq:EnsRe2_Final}, and~\eqref{eq:EnsImRe_Final} into~\eqref{eq:variancePhi} yields the explicit expression for the variance of correlation phases: 
\begin{equation}
\sigma^2_\phi({r_\text{c}}) \approx  
 \sqrt{\pi}\frac{\bar{\sigma}_\text{A}}{L}  \left[B^{-2}_0({r_\text{c}}) - 1\right].   
\end{equation}